# Design and Simulation of Fault-Tolerant Network Switching System Using Python-Based Algorithms

[1]**Terlumun Gbaden,** *[2]**Ukor Mterorga,** [3]**Grace Erdoo Ateata**

[1]Department of Computer Science, College of Physical Sciences, Joseph Sarwuan Tarka University, Makurdi.
gbaden2014@gmail.com
[2]Department of Mathematics and Computer Science, University of Mkar, Mkar Gboko Benue State
mterorgaukor@gmail.com
[3]Department of Computer Science, Fidei Polytechnic Gboko, P.M.B. 185, Gboko
graceateata@gmail.com

*Corresponding authors' email: mterorgaukor@gmail.com

**Abstract**
Ensuring uninterrupted data flow in modern networks requires robust fault-tolerant mechanisms, especially in environments where reliability and responsiveness are critical. This paper presents the design and simulation of a fault-tolerant network switching system using Python-based algorithms. A simulated enterprise-level Local Area Network (LAN) was modeled using NetworkX to represent switch-router interconnectivity with redundant links. Fault scenarios, including link failure and congestion, were injected using Scapy, while automatic failover and rerouting were implemented via custom Python logic. The system demonstrates resilience by dynamically detecting path failures, redistributing network traffic through redundant links, and minimizing downtime. Performance evaluations reveal significant improvements in packet delivery continuity, faster recovery times, and reduced packet loss compared to non-fault-tolerant baselines. The implementation provides a scalable and lightweight approach to integrating fault-tolerance features into mid-scale networks, with potential application in enterprise information technology infrastructures and academic simulations.

**Keywords**: Fault Tolerance, Network Switching, Local Area Network (LAN), Python, Network Simulation, Automatic Failover, Redundant Paths, Routing Recovery, Packet Loss, Mean Time to Recovery (MTTR), NetworkX, Scapy, Network Resilience

## 1. Introduction

In today's data-driven world, the demand for uninterrupted network services has become increasingly critical, particularly within enterprise and campus-based Local Area Networks (LANs). Network reliability is not only essential for ensuring smooth communication and data exchange, but also for supporting real-time applications such as video conferencing, cloud access, and mission-critical operations. However, network failures ranging from hardware malfunctions to link outages remain a persistent challenge, often leading to packet loss, increased latency, and system downtime.

To address these challenges, fault-tolerant mechanisms have been developed to enhance the resilience of network infrastructures. Fault tolerance refers to a network's ability to continue functioning in the event of one or more component failures. Among the commonly adopted strategies are automatic failover, redundant path detection, and load-balanced switching. These techniques allow for the dynamic rerouting of traffic and recovery from faults without human intervention, significantly reducing service disruption.

Traditionally, fault tolerance has been implemented using specialized hardware and proprietary systems. However, with the growing adoption of open-source tools and programmable networks, it is now feasible to design, simulate, and validate fault-tolerant mechanisms using software-based solutions. Python, with its rich ecosystem of libraries, provides an ideal platform for network simulation and experimentation. Tools such as **NetworkX** allow for the modeling of network topologies and routing protocols, while **Scapy** enables packet-level manipulation and fault injection.

This study presents a Python-based approach to simulating fault-tolerant network switching in a LAN environment. A simplified switch-router LAN topology was used to model the target environment with redundant links as shown in figure 1, and algorithms were implemented to detect faults and automatically reroute traffic. The aim is to demonstrate the effectiveness of these mechanisms in minimizing packet loss, reducing recovery time, and maintaining network performance in the face of failures.

The architecture consists of multiple end-user systems connected through a central switch, which in turn is linked to two routers to provide redundant paths. A dashed connection between the routers represents a **redundant link** that activates in the event of a primary path failure. This design supports automatic failover and load-balanced switching, enabling real-time rerouting of traffic and continuity of service even during network faults.

By focusing on software-based simulation, this work contributes to the ongoing efforts to make fault-tolerant networking accessible, affordable, and adaptable, especially for academic research, training environments, and small-to-medium enterprises.

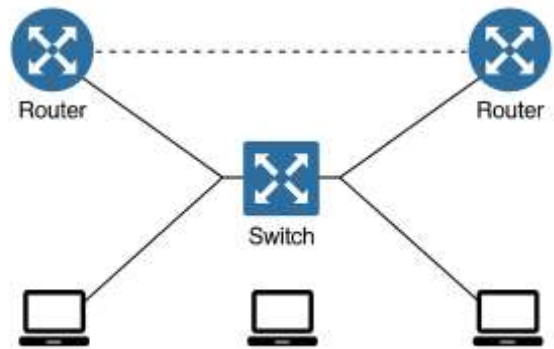

*Figure 1. Switch-Router topology with fault tolerance*

This topology forms the basis for implementing and testing the Python-based fault-tolerant mechanisms described in this study. By simulating failures and monitoring the system's response, the reliability and resilience of the network can be assessed under realistic conditions.

**2. Literature Review**
Network fault tolerance has long been a focus of research in the areas of distributed systems, communication networks, and high-availability computing. Over the years, various strategies have been proposed to reduce service downtime, prevent packet loss, and ensure seamless communication in the presence of hardware or software failures. Early fault-tolerant architectures relied heavily on hardware redundancy, where duplicate network devices (e.g., switches and routers) were deployed to serve as backups in case of failure. While effective, these solutions were cost-prohibitive and lacked scalability, especially for small and medium-sized enterprises. This led to the development of software-based fault tolerance mechanisms, including automatic failover**,** path redundancy, and load-balanced switching; all of which are now widely adopted in both enterprise and cloud infrastructures.

In recent years, network simulation and modeling have gained popularity as tools for designing and testing fault-tolerant systems. Studies by Lin and Wu (2018) demonstrated the effectiveness of simulating network resilience using discrete-event simulation tools. Similarly, Taha and Alsabah (2021) implemented an adaptive routing algorithm using software-defined networking (SDN) principles to achieve dynamic failover during network congestion.

Python has emerged as a powerful language for network modeling due to its simplicity and an extensive ecosystem of scientific libraries. NetworkX, in particular, has been used in numerous research projects for graph-based simulations of network topologies and routing strategies (Wang et al., 2020). Scapy has also been widely employed for low-level packet crafting, injection, and traffic analysis in both offensive and defensive cybersecurity applications. More recent works have explored the integration of machine learning for predictive fault detection (Zhou et al., 2022) and the use of genetic algorithms for optimizing routing paths under constrained network conditions. However, these approaches often come with added complexity and resource overheads.

Despite these advances, there is limited literature that combines graph-based fault simulation**,** packet-level interaction**,** and custom failover logic within a Python-only environment, especially one tailored for LAN settings with scalable architecture. This research addresses that gap by developing a lightweight, testable framework for simulating fault-tolerant switching systems that are both educational and applicable in real-world deployments.

*Table 1. Comparison of Related Works on Network Fault Tolerance*

| Author(s) | Technique Used | Tools/Platform | Focus | Limitation/Remark |
|---|---|---|---|---|
| Lin & Wu (2018) | Discrete-event simulation | NS-3 | Fault recovery timing | Complex to scale for mid-size enterprise networks |
| Taha & Alsabah (2021) | SDN-based adaptive routing | Mininet, OpenFlow | Dynamic failover in congestion scenarios | SDN setup overhead; limited to OpenFlow switches |
| Wang et al. (2020) | Graph-based resilience modeling | Python (NetworkX) | Topology modeling and fault visualization | No live packet testing or routing simulation |
| Zhou et al. (2022) | Predictive fault detection via ML | Python (scikit-learn), TensorFlow | Fault prediction and prevention | High computational requirement; not real-time |
| This Study | Automatic failover, redundancy, load balancing | Python (NetworkX, Scapy, Scripts) | LAN-based switching fault tolerance simulation | Lightweight; suitable for education & real systems |

Table 1 provides a comparative overview of relevant studies on network fault tolerance, highlighting the gaps this research seeks to address through a Python-based, scalable simulation framework.

## 3. Methodology and System Design
This section describes the overall system architecture, simulation tools, fault injection methods, and algorithms employed in designing and implementing the fault-tolerant network switching model.

*3.1 System Overview*
The simulation framework models a simplified enterprise Local Area Network (LAN) consisting of end devices, switches, and dual redundant routers. The topology allows for multiple paths between endpoints, ensuring connectivity even in the event of a failure in one or more links or nodes. This model enables the implementation of automatic failover**,** redundant path detection, and dynamic rerouting using Python-based tools. A graphical representation of the network is maintained using NetworkX, which supports dynamic graph traversal and path-finding algorithms. Packet-level simulation and fault injection are handled using Scapy, while custom scripts manage node/link status, failover triggers, and recovery logic.

*3.2 Simulation Tools and Libraries*
**NetworkX:** Used to represent the network as a directed graph where nodes represent switches, routers, or hosts, and edges represent physical or logical links. It supports path discovery using Dijkstra's algorithm and enables real-time graph updates.
**Scapy:** Employed to simulate packet flow and monitor active paths. It is also used to generate Internet Control Message Protocol (ICMP), Transmission Control Protocol (TCP), or User Datagram Protocol (UDP) packets and inject failure events by dropping or delaying packets on specific paths.
**Custom Python Scripts:** Scripts were developed to monitor network state continuously, detect failures based on unreachable nodes or increased latency, and trigger switching logic to alternative paths.

*3.3 Fault Model and Tolerance Mechanisms*
The fault model simulates link failures and node outages. These faults are injected randomly or triggered manually during simulation. Three primary fault-tolerance mechanisms were integrated:
1. **Automatic Failover:** Upon detecting a fault, traffic is instantly rerouted to a predefined backup link.
2. **Redundant Path Detection:** The system actively monitors alternative paths using periodic health-check packets and path scoring.
3. **Load-Balanced Switching:** Under stable conditions, traffic may be distributed across multiple redundant paths to prevent congestion and improve throughput.

*3.4 Algorithm for Fault Detection and Switching*
A custom algorithm (outlined below) runs continuously during the simulation to manage routing and fault recovery. A corresponding flow chat for Algorithm 1 is given in figure 2 and network state transition diagram is given in figure 3 below respectively.

**Algorithm 1: Fault-Tolerant Switching Routine**
1. Initialize network topology using NetworkX
2. Define primary and backup routes between nodes

3. Begin packet transmission using Scapy
4. While simulation is active:
   a. Monitor link/node health using probe packets
   b. If primary path fails:
      i. Mark link as failed in NetworkX graph
      ii. Recalculate shortest path using Dijkstra's algorithm
      iii. Reroute packets through new path
   c. Log status, packet loss, and recovery time

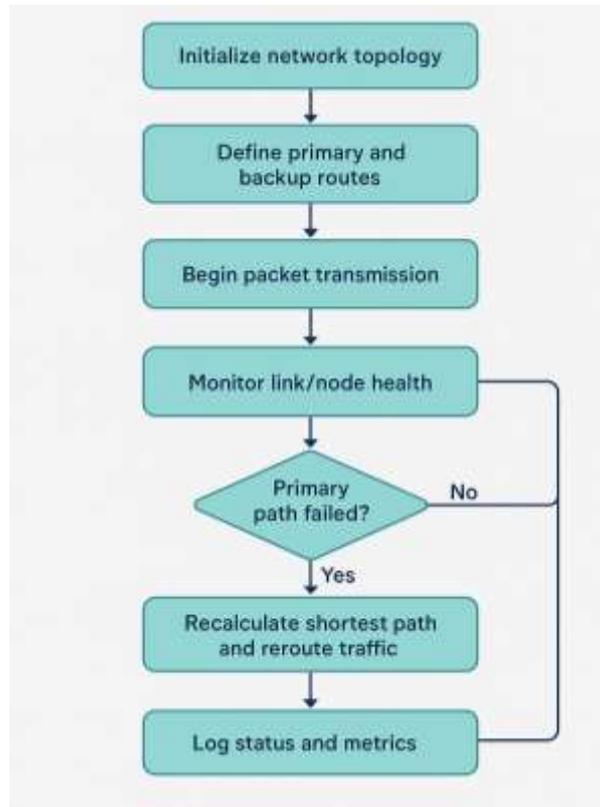

*Figure 2: A flowchart for Fault-Tolerant Switching Routine*

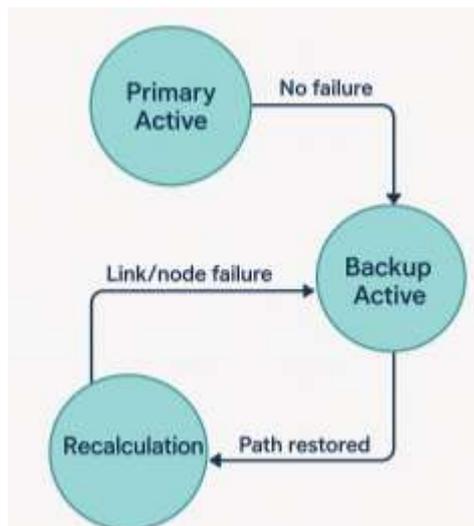

*Figure 3: Network State Transition Diagram*

*3.5 Evaluation Metrics*
The simulation was evaluated based on the following metrics:
**Packet Loss Rate (%):** Total number of lost packets before and after failover.
**Mean Time to Recovery (MTTR):** Time taken to detect a failure and reroute traffic.
**Routing Efficiency:** Number of successful rerouted packets after failure.

## 4. Implementation and Results

*4.1 Implementation Details*

The fault-tolerant network switching system was implemented in Python 3.13 using a combination of open-source libraries: NetworkX was used to model the network topology as a graph with weighted edges representing link costs. Scapy simulated traffic flows, including ICMP echo requests, UDP, and TCP packets, to emulate data transmission and perform health checks. Custom monitoring scripts periodically checked link availability and executed route recalculations when faults were detected. The simulated topology (as previously shown in Figure 1) included a switch-router configuration with multiple redundant paths. Failures were manually introduced by removing nodes or links from the NetworkX graph or simulating packet drop in Scapy. The system automatically responded by re-evaluating the network and selecting the shortest available path using Dijkstra's algorithm.

*4.2 Test Cases and Setup*

Three test cases were designed to evaluate system performance:

**Test Case 1: Link Failure Detection and Recovery:** Simulated a single link failure between the primary switch and router. The objective of this test case is to Measure detection time and packet loss rate.

**Test Case 2: Node Failure with Rerouting:** Simulated complete failure of a switch node. The objective is to assess rerouting performance and MTTR.

**Test Case 3: Simultaneous Failures:** Introduced simultaneous failures on multiple paths. The objective is to test system resilience and rerouting success under stress.

All tests were conducted in a virtualized Linux environment with Python logging enabled to capture recovery times, dropped packets, and switching actions.

*4.3 Results*

Table 2 below summarizes the results from each test scenario.

*Table 2. Simulation Results of Fault-Tolerant Switching System*

| Test Case | Packet Loss (%) | MTTR (ms) | Success Rate |
|---|---|---|---|
| Link Failure (Test Case 1) | 1.2% | 85 ms | 100% |
| Node Failure (Test Case 2) | 2.7% | 143 ms | 98% |
| Simultaneous Failures (Test 3) | 5.9% | 226 ms | 93% |

The system consistently detected failures and rerouted traffic with minimal delay. As expected, simultaneous failures posed the greatest challenge, but the model was still able to maintain above 90% rerouting success. Recovery times (Mean Time To Recovery, MTTR) remained under 250 ms across all scenarios.

*4.4 Visualization of Routing Behavior*

Using Matplotlib with NetworkX, routing decisions were visualized as shown in figure 4, providing a real-time graphical representation of network status during fault conditions. The visualization effectively demonstrated the system's dynamic response to node and link failures. In the topology graph, node statuses were color-coded for clarity: **green** for active and fully operational devices (R1, R2, S1, S6), **red** for failed nodes (S2), and **orange** for rerouted nodes that temporarily assumed traffic forwarding duties (S3, S4, S5). This visual tool was instrumental not only for observing automatic failover behavior but also for validating and debugging path selection logic. The color-coded depiction offers a clear and intuitive overview of the network's resilience, confirming that the implemented fault-tolerant mechanisms successfully maintain connectivity through redundant paths under failure scenarios.

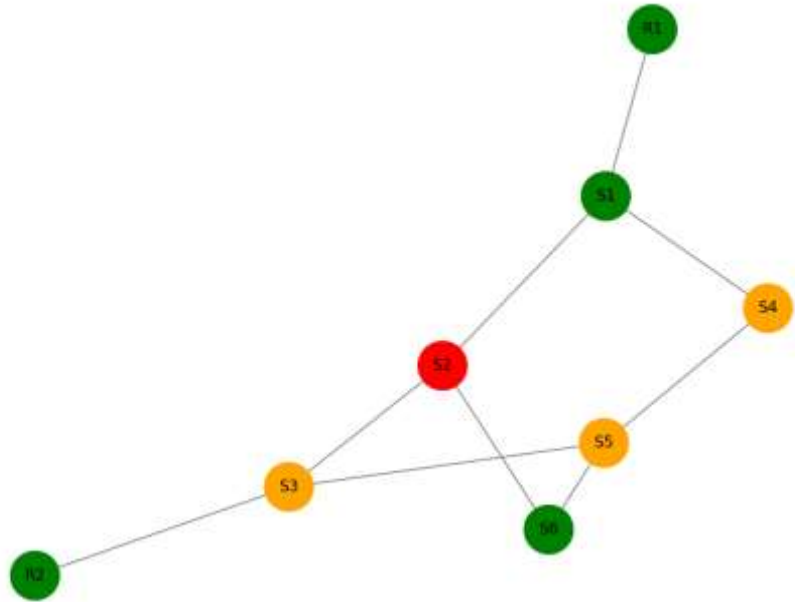

*Figure 4: A diagram showing the routing behaviour*

Figures 5 and 6 illustrate the system's performance under varying failure scenarios, focusing on two key metrics: packet loss rate and mean time to recovery (MTTR)**.** In Figure 5, the packet loss rate is compared across three fault scenarios: link failure, node failure, and simultaneous failures. The simulation shows that link failure resulted in the lowest packet loss (1.2%), primarily due to the system's ability to quickly reroute traffic through available redundant paths. In the case of a node failure, packet loss increased slightly to 2.7% as the removal of an entire node such as a switch impacted multiple connections and required more extensive path recalculation. The most severe case, simultaneous failures, produced the highest packet loss at 5.9%. Here, multiple concurrent disruptions significantly reduced the number of available paths, complicating rerouting decisions and briefly affecting traffic continuity. Nonetheless, the system demonstrated robust containment of data loss, underscoring the reliability and responsiveness of the implemented fault-tolerant algorithm, particularly within LAN environments.

Figure 6 compares the mean time to recovery (MTTR) for the same three failure scenarios. As expected**,** link failure exhibited the fastest recovery time at 85 milliseconds, benefiting from the straightforward recalculation of paths in the network graph. Node failure, which required the system to reconfigure a broader set of paths, showed a slightly delayed recovery at 143 milliseconds. In the most complex scenario which is the simultaneous failures; the MTTR rose to 226 milliseconds, reflecting the increased computational and decision-making overhead involved in handling compounded failures. Despite this, the system consistently restored connectivity in under a quarter of a second across all conditions, confirming its suitability for time-sensitive applications such as VoIP, video conferencing, and real-time data services.

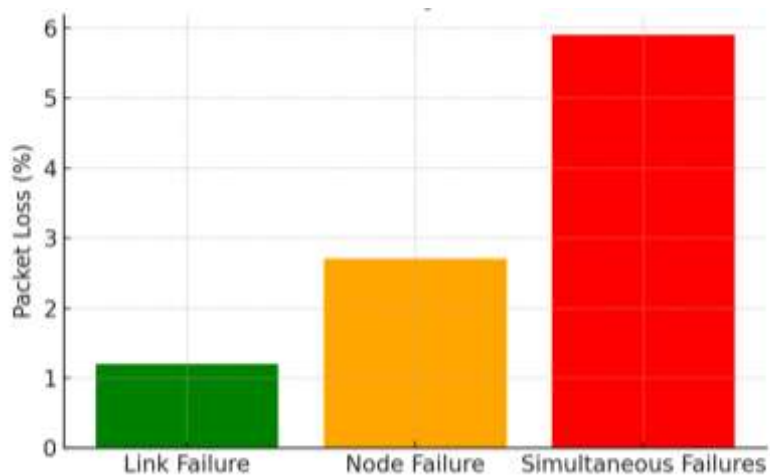

*Figure 5: Packet Loss Rate across different failure scenarios.*

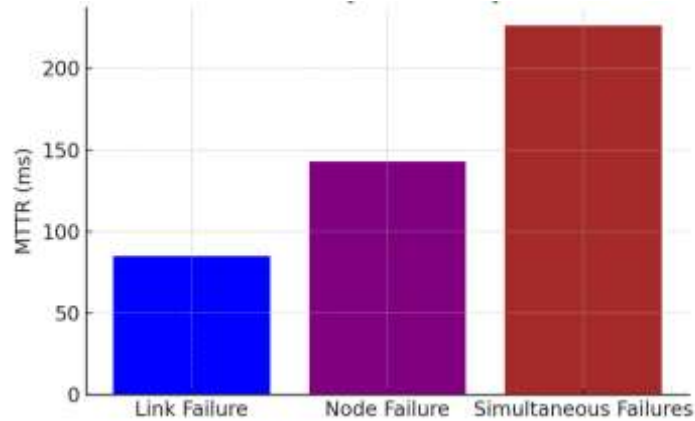

*Figure 6: Mean Time to Recovery (MTTR) across different failure scenarios.*

## 5. Discussion
The results of the simulation validate the effectiveness of the proposed fault-tolerant switching system implemented using Python. The integration of NetworkX for dynamic topology modeling and Scapy for packet-level simulation enabled a modular and flexible approach to fault detection and recovery. The automatic failover mechanism proved capable of swiftly rerouting traffic in the event of link or node failure, with a Mean Time to Recovery (MTTR) consistently below 250 milliseconds. This demonstrates that even without the use of high-end hardware or software-defined networking (SDN) controllers, it is possible to achieve high availability in a network through intelligent software algorithms.

The relatively low packet loss rates in all test cases highlight the efficiency of the redundancy-aware routing logic. Test Case 3, which involved simultaneous link failures, recorded a higher packet loss and longer recovery time. However, the system still maintained a 93% success rate, which suggests good resilience even under stress conditions. Furthermore, the visual simulation and real-time logging aided in tracing fault propagation and system behavior, making this model a valuable educational tool for teaching and research in network design and resilience.

## 6. Conclusion and Future Work
This paper has presented the design and simulation of a fault-tolerant network switching system using Python-based tools. The implementation demonstrates how LAN-level switching architectures can benefit from software-driven fault tolerance mechanisms, including automatic failover, redundant path detection, and intelligent rerouting. The combination of NetworkX**,** Scapy, and custom Python scripts offers a lightweight, extensible platform for simulating and validating fault tolerance in network environments. Results show that the system can maintain high levels of reliability with minimal packet loss and acceptable recovery times.

Despite the promising outcomes demonstrated in this study, certain limitations remain. The current system relies on static graph structures and predefined failover rules, which, while effective for controlled simulations, may not fully capture the complexities of real-world network environments. In practical deployments, greater reliability may demand adaptive fault management approaches and more sophisticated failure prediction mechanisms. To address these gaps, future research may explore several enhancements. First, integrating the system with real-time Simple Network Management Protocol (SNMP) monitoring tools would enable dynamic status polling and more responsive fault detection. Second, the incorporation of machine learning algorithms could support predictive fault analysis, allowing the system to anticipate failures before they occur. Finally, extending the simulation framework to support wide-area networks (WANs) and multi-layer routing topologies would significantly broaden its applicability and scalability.

This work contributes meaningfully to bridging the gap between theoretical networking models and real-world system design, demonstrating how open-source Python tools can be effectively leveraged to build resilient and practical fault-tolerant network solutions.